\documentclass[preprint]{aastex}

\usepackage{color,amsmath}
\newcommand\Msun{\mbox{$M_\sun$}}
\newcommand\kmps{\mbox{km\,s${}^{-1}$}}

\shorttitle{IRC+10216 CENTRAL CORE}
\shortauthors{KIM, Lee, Mauron, \& Chu}

\begin{document}
\title{HST IMAGES REVEAL DRAMATIC CHANGES IN THE CORE OF IRC+10216}
\author{Hyosun Kim\altaffilmark{1,2,3}}
\author{Ho-Gyu Lee\altaffilmark{1}}
\author{Nicolas Mauron\altaffilmark{4}}
\author{You-Hua Chu\altaffilmark{2}}
\altaffiltext{1}{Korea Astronomy and Space Science Institute 776, 
  Daedeokdae-ro, Yuseong-gu, Daejeon, Republic of Korea 305-348}
\altaffiltext{2}{Academia Sinica Institute of Astronomy and Astrophysics, 
  P.O. Box 23-141, Taipei 10617, Taiwan; hkim@asiaa.sinica.edu.tw}
\altaffiltext{3}{EACOA fellow}
\altaffiltext{4}{Universite de Montpellier, Batiment 13 CC072, Place 
Bataillon, 34095 Montpellier, France}

\begin{abstract}
IRC+10216 is the nearest carbon star with a very high mass-loss rate. The 
existence of a binary companion has been hinted by indirect observational 
evidence, such as the bipolar morphology of its nebula and a spiral-like 
pattern in its circumstellar material; however, to date, no companion has 
been identified. 
We have examined archival \emph{Hubble Space Telescope} images of 
IRC+10216, and find that the images taken in 2011 exhibit dramatic 
changes in its innermost region from those taken at earlier epochs. 
The scattered light is more spread out in 2011. After proper motion 
correction, the brightest peak in 2011 is close to, but not coincident 
with, the dominant peak in previous epochs. A fainter point-like object 
was revealed at $\sim$ 0\farcs5 from this brightest peak. 
We suggest that these changes at the core of IRC+10216 are caused by 
dissipation of intervening circumstellar dust, as indicated by the 
brightening trend in the lightcurve extracted from the Catalina 
photometric survey. We tentatively identify the brightest peak in 
2011 as the primary star of IRC+10216 and the fainter point-like 
source as a companion. The cause of non-detections of the companion 
candidate in earlier epochs is uncertain. These identifications 
need to be verified by monitoring of the core of IRC+10216 at high 
resolution in the future.
\end{abstract}


\section{INTRODUCTION}\label{sec:int}

Direct detection of binary companions of asymptotic giant branch 
(AGB) stars is challenging because the companions are overwhelmed 
by the luminous AGB stars. Moreover, companions of very dusty 
``infrared'' AGB stars can be totally obscured by the circumstellar 
envelopes (CSEs). To date, binary companions have been detected for 
only a small number of AGB stars with low mass-loss rates through 
composite spectra or lightcurves \citep{jor03}. There are indications 
that a significant fraction of AGB stars have binary companions, 
such as elliptical CSEs \citep{mau13} as expected from binary AGB 
stars \citep{hug09} and spiral-shell patterns in the CSEs \citep
{mau06,may11,mae12,kim13,may14,cer14,dec15} as modeled by AGB winds 
under the gravitational influence of binary companions \citep
{the93,sok94,mas99,kim12a,kim12b,kim12c,kim13}. 
 
Since its discovery \citep{bec69}, IRC+10216, the nearest carbon 
star, has been the subject of multi-wavelength 
observations aimed at detecting new molecules or studying the 
circumstellar dust and chemistry \citep[e.g.,][]{gue77,jur83,
gla96,gue11}. The overall shape of its CSE, detected up to radius 
$\sim200\arcsec$, is fairly spherical. Due to its proximity \citep
[distance $\sim130$\,pc,][]{gro98,gro12}, the envelope has been 
resolved, exhibiting multiple shells with noticeably different 
geometric centers \citep{cra87,mau99,mau00,fon03,dec11}. 
The dynamical timescale of shell separation, 200--800 years, is far 
longer than the pulsation period of IRC+10216 \citep[$\sim$ 649 days,]
[]{leb92} and significantly shorter than the periods between thermal 
pulses (of order of $10^4$ years, \citealp{blo95}). The shell 
geometry must have been caused by other mechanisms, possibly binary 
interactions as considered previously \citep[e.g.,][]{mau99,dec11}.
Indeed, the shell center offsets can be reproduced by a spiral-shell 
model of IRC+10216 with enhanced mass loss when a companion approaches
periastron \citep[e.g.,][]{cer14}. 
Recent Atacama Large Millimeter/submillimeter Array (ALMA) observations 
show that the morphology and kinematics of its inner parts are consistent 
with those expected from binary interactions \citep{dec15}. 

High-resolution optical images of IRC+10216 taken in the years around 
2000 revealed a bipolar nebula $\sim2\arcsec$--$3\arcsec$ in size, 
with the northeast lobe significantly fainter than the southwest lobe 
\citep[e.g.,][]{han98,ski98,mau00,lea06}. This nebula was further 
modeled as an inclined bi-conical structure \citep{dyc87,han98}. Upon 
examination of all archival \emph{Hubble Space Telescope} (\emph{HST}) 
images of IRC+10216's core, we find that in 2011 the aforementioned 
bipolar nebula had disappeared and a different set of peaks appeared.
We identify the brightest peak and a fainter 
point-like source as the central star and its companion. In this paper, 
we report our analysis of archival data of IRC+10216 and discuss its 
physical nature.

\section{Archival Data}\label{sec:obs}
\subsection{\emph{Hubble Space Telescope} Images}\label{sec:obs1}

Three images of IRC+10216 taken with the F606W filter were retrieved 
from the \emph{HST} archives: Epoch 1 on 1998 March 30 with the WFPC2 
camera (PID 6856; P.I. John Trauger) for a total integration time of 
$t=4000$\,s; Epoch 2 on 2001 January 7 with the WFPC2 and $t=600$\,s 
(PID 8601; P.I. Patrick Seitzer), and Epoch 3 on 2011 June 4 with 
the WFC3 camera and $t=5407$\,s (PID 12205; P.I. Toshiya Ueta). 
The WFPC2 and WFC3 fields are about 2\farcm5$\times$2\farcm5 and 
3\arcmin$\times$3\arcmin, and the pixel sizes are 0\farcs1 and 
0\farcs04, respectively. The three F606W images are displayed in 
Figure\,\ref{fig:hst}. It can be seen at once that a very drastic 
change of structure has occurred between Epoch 2 and 3. A F814W 
image taken at Epoch 3 (Figure\,\ref{fig:hst}(d)) from the same 
program is saturated at the brightest peak but shows an, otherwise, 
almost identical morphology to the F606W image. 

To compare these images, an astrometric analysis was performed using the 
CCMAP package of IRAF \citep{tod93}. Three well-detected stars located 
within 40\arcsec\ from IRC+10216 and identified in the Sloan Digital Sky 
Survey (SDSS) catalog \citep{aih11} were chosen for reference.  Only one 
of these stars is well-detected in the shallow Epoch 2 image; therefore, 
we use two compact galaxies and one star to bootstrap the astrometry from 
the Epoch 1 and 3 images. The combined uncertainty in the astrometric 
alignment between each pair of images is $\le$0\farcs08. 

\subsection{Photometric Monitoring Data from the Catalina Sky Survey}
\label{sec:obs2}

We have used the Catalina Sky Survey \citep{dra14} data to investigate 
the variability of IRC+10216. The Catalina lightcurve of IRC+10216, 
shown in Figure\,\ref{fig:lc3}(a), covers 2005--2014. 
As the Catalina Sky Survey used unfiltered images to maximize throughput, 
these photometric measurements correspond to passbands depending on
a star's spectral energy distribution (SED) and the CCD responses. 
IRC+10216 is a very red object with possible phase-dependent color variations. 
Therefore, we compared the Catalina photometry with contemporaneous SDSS 
data (from 2006 January 6) in different passbands, and found that the SDSS 
$i$ band had a similar value \citep[indicated by a filled cyan circle 
in Figure\,\ref{fig:lc3}(a);][]{ahn12}. The SDSS $r$ and $z$ band magnitudes 
were off by $+3.2$ and $-2.4$ mag, respectively. 

We searched for similar monitoring data from earlier epochs, and found 
two with extensive coverage and both were in infrared wavelengths: $JHKL'M$ 
for 1985--1989 \citep{leb92} and $JHKLM$ for 2000--2008 \citep{she11}. 
The latter overlapped the Catalina monitoring for 3--4 years. 
These infrared lightcurves are shown in Figure\,\ref{fig:lc3}(b)--(c).

\section{Optical and Near-Infrared Variations}\label{sec:res}
\subsection{Morphological Variations in the Core}\label{sec:hst}

The \emph{HST} F606W images from the three epochs are shown in 
Figure\,\ref{fig:hst}. The 1998 and 2001 images are qualitatively 
similar, exhibiting a bright southwest lobe and a fainter northeast 
lobe, previously identified as a bipolar structure \citep[e.g.,]
[]{mau00}. The 2011 image shows a dramatically different morphology: 
the brightest peak appears to be at a different location, and the 
region within $1\arcsec$ from the peak has a more complex structure, 
including a point-like source at 0\farcs5 east of the peak. 
There is no trace of the previously identified bipolar structure. 
Using a 6\arcsec-radius source aperture and a background annulus of 
radii 14\arcsec--16\arcsec, we have measured the \emph{HST} F606W 
magnitudes of IRC+10216 to be 17.29, 17.15, and 16.93 mag in 1998, 
2001, and 2011, respectively.

\begin{figure*} 
  \plotone{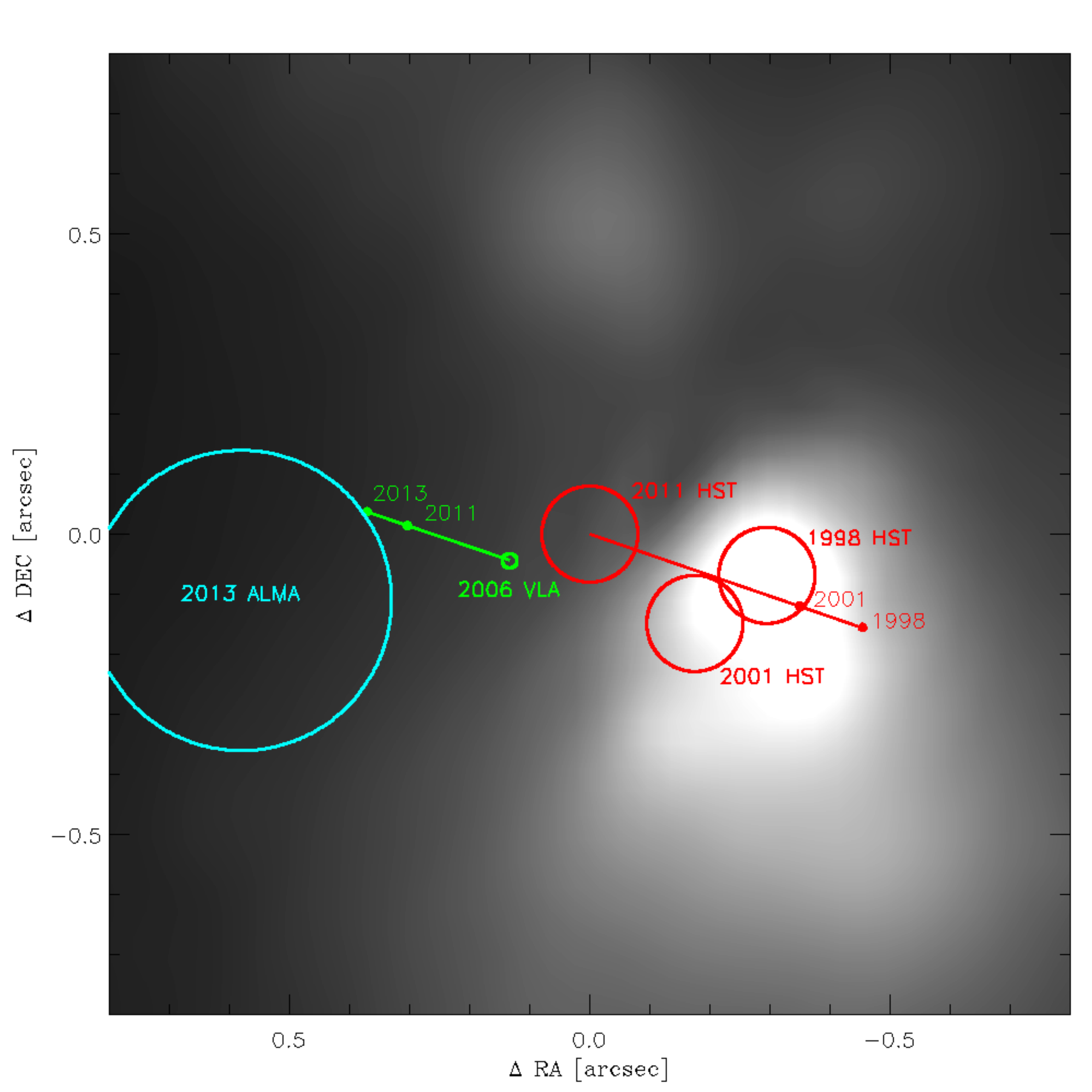}
  \caption{\label{fig:pro}
    Relative positions of brightest peaks of the \emph{HST} F606W images at 
    epoch 1998, 2001, and 2011 (red circles), and radio continuum emission 
    from the VLA at epoch 2006 \citep[green circle,][]{men06} and from the 
    ALMA at epoch 2013 \citep[cyan circle,][]{dec15} on the background 1998 
    \emph{HST} image. The radii of circles represent the uncertainties of 
    astrometry for the radio data and of alignment between the \emph{HST} 
    images. Linear lines and dots show the proper motions of the optical 
    2011 HST (red) and radio 2006 VLA (green) peaks, and the expected 
    positions at epochs written next to the dots. 
  }
\end{figure*}

Images from the three epochs have been aligned astrometrically 
with respect to background stars and galaxies; however, to properly 
determine the differences among the images, the proper motion of 
IRC+10216 has to be considered.  Using radio continuum observations
made with the Very Large Array (VLA), the proper motion of IRC+10216 
has been reported to be $35\pm1$ mas yr$^{-1}$ in RA and $12\pm1$ 
mas yr$^{-1}$ in Dec \citep{men12}. For this proper motion, the 
brightest peak in 2011 would correspond to positions close to, 
albeit not exactly on, the peaks in 1998 and 2001. The offsets 
(both about 0\farcs18) are small, but larger than the uncertainty 
in the differential astrometry ($\le$0\farcs08). The observed peak 
positions from 1998, 2001, and 2011 and the expected 1998 and 2001 
positions of the 2011 peak are plotted in Figure\,\ref{fig:pro}.  

It is interesting to compare the 2011 peak position with the central 
position of IRC+10216 reported from high-resolution radio observations. 
The radio continuum emission of IRC+10216 originates from its radio 
photosphere \citep{men06}, and its J2000 coordinates observed with the
VLA in Epoch 2006.16 are 09$^{\rm h}$47$^{\rm m}$57$^{\rm s}$.4255, 
$+13\arcdeg16\arcmin43\farcs815$ \citep{men12}. The J2000 coordinates 
of IRC+10216 derived from ALMA observations at a higher radio frequency 
in the Epoch 2012.92 are 09$^{\rm h}$47$^{\rm m}$57$^{\rm s}$.4553, 
$+13\arcdeg16\arcmin$43\farcs749, different from the proper-motion-corrected 
VLA position by $\sim$0\farcs26, close to the error limit \citep{dec15}. 
Comparing these positions with the optical position of the 2011 peak 
(09$^{\rm h}$47$^{\rm m}$57$^{\rm s}$.4189, $+13\arcdeg16\arcmin43\farcs859$;
see Figure\,\ref{fig:pro}), we see a 0\farcs3 offset between the optical 
peak in 2011 and the VLA radio peak expected in 2011. Considering that 
the absolute astrometry of \emph{HST} can reach 0\farcs1--0\farcs3 
when a large number of reference stars are available \citep{koe06}, 
the uncertainty of our absolute astrometry of IRC+10216 based on only 
3 reference stars must be at least 0\farcs3. Therefore, we suggest 
that the radio peak and the 2011 optical peak are coincident within 
the uncertainty of the \emph{HST} astrometry. 

\subsection{Trends in the Lightcurves of IRC+10216}\label{sec:lcs}

The Catalina lightcurve of IRC+10216 in Figure\,\ref{fig:lc3}(a) 
shows a gradual trend of brightening from 2005 to 2014, in 
addition to the stellar pulsation of the carbon star. To quantify 
the brightening, we fit the lightcurve in Figure\,\ref{fig:lc3}(a) 
with a sinusoidal term and a linear term: 
\begin{equation}\label{eqn:fit}
  m (t) = (A_0/2)\cos(2\pi (t-A_1)/A_2) + (A_3+A_4t),
\end{equation}
where $m(t)$ is the observed magnitude at epoch $t$ and $A_0$--$A_4$ are 
constants. The best-fit model has $A_0=1.8\pm0.02$ mag for the amplitude 
of the sinusoidal variations; $A_2=640\pm0.9$ days for the period of the 
sinusoidal variations; and $A_4=-0.16\pm0.004$ mag yr$^{-1}$ for the 
brightening rate in the Catalina photometry. The derived pulsation 
period of the carbon star is consistent with the range of values from 
infrared bands (649 days, \citealp{leb92}; 630 days, \citealp{men12}).
No similar monitoring records at earlier epochs in unfiltered or $i$ band 
are available to show the onset of the brightening trend. The \emph{HST} 
F606W measurements span 14 years, but do not provide sufficient data points 
to constrain the parameters in Equation (\ref{eqn:fit}).

IRC+10216 was monitored in the infrared bands in 1985--1989 \citep{leb92} 
and 2000--2008 \citep{she11}. 
These lightcurves are plotted in Figure\,\ref{fig:lc3}(b) and (c), and 
fitted by Equation (\ref{eqn:fit}). We find that the $J$, $H$, and $K$ 
lightcurves in the 1980s show systematic brightening trends at rates of 
$A_4=-0.3$, $-0.2$, and $-0.1$ mag yr$^{-1}$, but at longer wavelengths 
the $L$ and $M$ lightcurves have nearly null brightening rates. 
Interestingly, the brightening trends in the near-infrared wavelengths 
vanished in the 2000s, when all observed infrared magnitudes stayed at 
a constant level with smaller amplitudes of sinusoidal variations (see 
Figure\,\ref{fig:lc3}(c)). The $H$, $K$, $L$, and $M$ magnitudes seem 
to become fainter by 0.3--0.7 mag from epoch 1989.5 to epoch 1999.9.

The lightcurves of IRC+10216 are clearly complex. The important features 
that need to be noted are: (1) there is the brightening trend of broad-band 
optical magnitudes in 2005--2014 and of near-infrared bands in 1985--1989;
(2) the near-infrared $JHK$ magnitudes vary around nearly constant levels
in 2000--2008, although the optical magnitudes show brightening in 2005-2008;
and (3) the mid-infrared bands of $L$ and $M$ show periodic variations around 
constant levels during all monitoring campaigns, regardless of the brightening
trends in other bands.

\section{DISCUSSION}\label{sec:dis}
\subsection{Nature of the Changes in the Core Region}

The innermost region in the CSE of IRC+10216 in the 2011 \emph{HST} 
image exhibits a strikingly different appearance from those of 1998 
and 2001 (Figure\,\ref{fig:hst}). 
This large modification of optical core structure has occurred in 
$\la$10 years; optical and infrared lightcurves have both shown 
brightening trends with a similar timescale. This timescale is much 
shorter than the stellar evolutionary timescale ($\sim10^3$--$10^4$ 
years) associated with thermal pulsing behavior and is not likely 
due to variation in the stellar luminosity. Since the dust envelope 
of IRC+10216 is optically thick, the stellar radiation is absorbed and 
reemitted mostly in infrared wavelengths; its SED peak is near $10\,\mu$m 
or beyond in mid-infrared \citep{lad10}. The lack of long-term variations 
in the mid-infrared $L$ and $M$ bands indicates that the stellar luminosity 
is steady. 

The timescale of $\la$10 years may be appropriate for photospheric or 
inner circumstellar changes. For example, a progressive rarefaction 
of the inner envelope allows penetration of stellar radiation, which can 
be detected in optical and near-infrared bands as a gradual brightening. 
The brightening trend could be also variable scattering due to 
changes in the geometry of the circumstellar material. 
Therefore, the brightening trends in the optical and near-infrared 
lightcurves likely imply the change of intervening circumstellar 
dusts in the $\la$10-year timescale, which may account for the recent 
changes of optical morphology in the core of IRC+10216.

\subsection{Optical Identification of Central Star}

The 2011 \emph{HST} F606W image of IRC+10216's core shows an elongated 
feature centered on the brightest peak. We have established in Section 
\ref{sec:hst} that this peak is coincident with the expected radio peak 
within the uncertainties of absolute astrometry; therefore, we identify 
this brightest peak as the central star of IRC+10216. 
The expected locations of this central star in 1998 and 2001 are both 
offset from the brightest peak by $\sim$ 0\farcs18 (or 23 AU). This offset 
is larger than the uncertainty in the relative astrometry ($\le$0\farcs08).
It is likely that the brightest peaks in 1998 and 2001 were both dust 
clump in close vicinity of the central star. 

A dust lane of IRC+10216, extending up to $4\arcsec$ in the east-west 
direction, is recently presented by \citet{jef14} in a linearly polarized 
optical intensity map. By placing the brightest peak in the \emph{HST} 
2011 image at their suggested location of central star in the dust lane, 
we have noticed that our elongated structure (PA $\sim-15\arcdeg$) is 
perpendicular to their dust lane (PA $\sim80\arcdeg$) and connects the 
brightest parts in the two lobes of polarized optical light. 
The width of our elongated structure is $\sim$ 10--20\,AU (or 5--10 
stellar radii with the photospheric radius of IRC+10216 $\sim$ 2 AU 
suggested by \citealt{men12}), which is similar to the radius of a 
dust formation zone or the inner radii used in the torus and ring 
models of \citet{jef14}. The elongated structure could be caused by 
either an illumination or a confined outflow through the 
central hole of a dense equatorial torus.

The elongated structure consists of two compact blobs on two opposite 
sides of our suggested central star. Both blobs are $\sim$ 0\farcs2 from 
the central star; assuming the same expansion speed as the outer envelope 
\citep[$\sim$ 15\,\kmps,][]{lou93}, they could be launched about 8 years 
ago. We also note that the elongated structure extends up to $\sim$ 0\farcs4, 
which corresponds to about 26 stellar radii. The elongated structure is 
slightly curved, thus shaped like a spiral or a point-symmetric structure. 

\subsection{Candidate for Companion of IRC+10216}

\begin{figure*} 
  \plotone{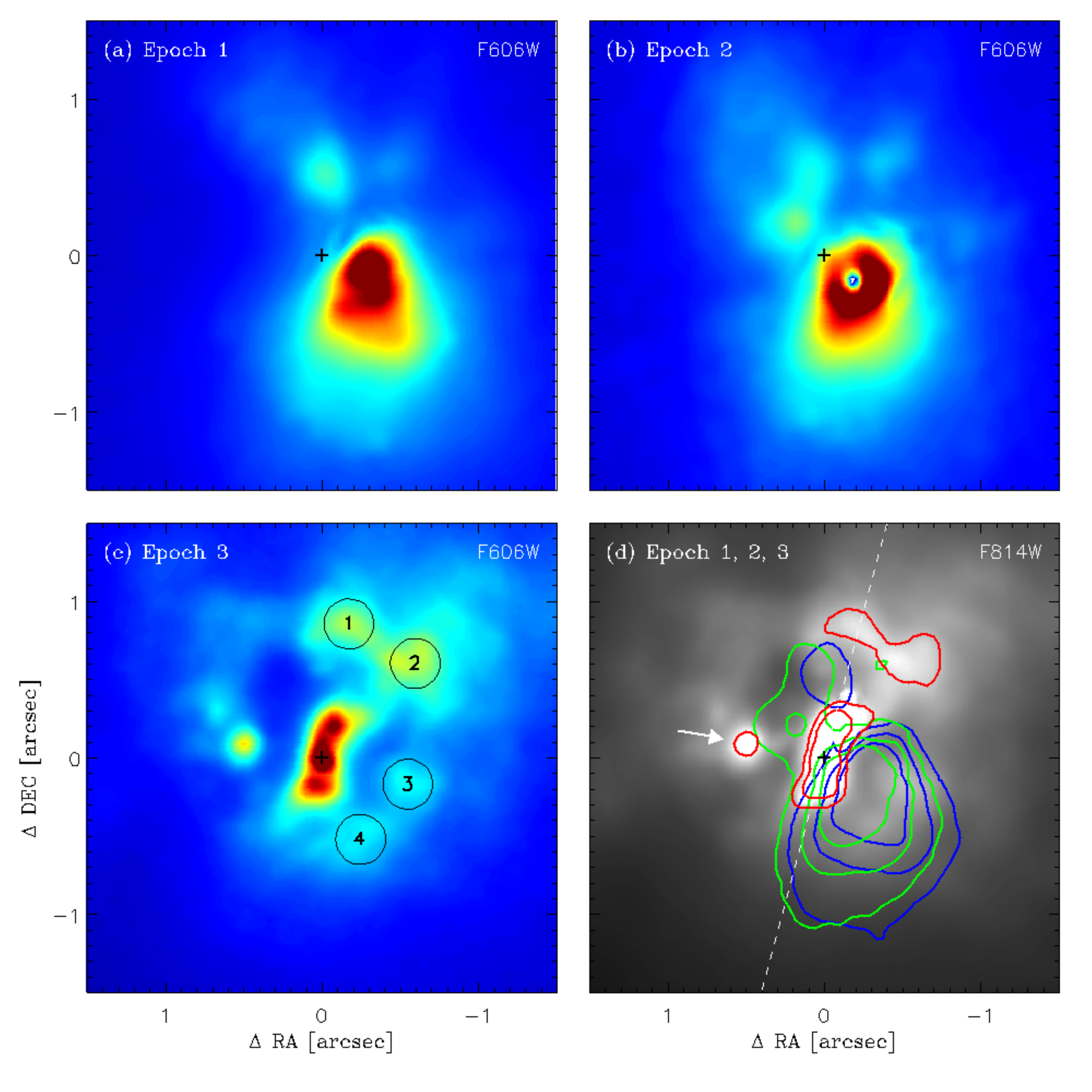}
  \caption{\label{fig:hst}
    \emph{HST} images of the core region of IRC+10216. 
    Panels (a)--(c) show images taken with the F606W filter from 
    Epoch 1 in 1998, Epoch 2 in 2001, and Epoch 3 in 2011, respectively. 
    Panel (d) shows an F814W image from Epoch 3 superposed with blue, 
    green, and red contours of F606W images from Epochs 1--3, respectively. 
    The cross at the center marks the position of the peak in 2011 image. 
    The four dust clumps marked with 0\farcs16-radius circles in panel (c) 
    are used for color comparisons with the candidate companion star marked 
    by an arrow in panel (d). The dashed line in panel (d) marks the PA of 
    $-15\arcdeg$ along the central elongated feature in Epoch 3. 
    Note that the peak of the image in panel (b) is saturated.
  }
\end{figure*}

\begin{figure*} 
  \plotone{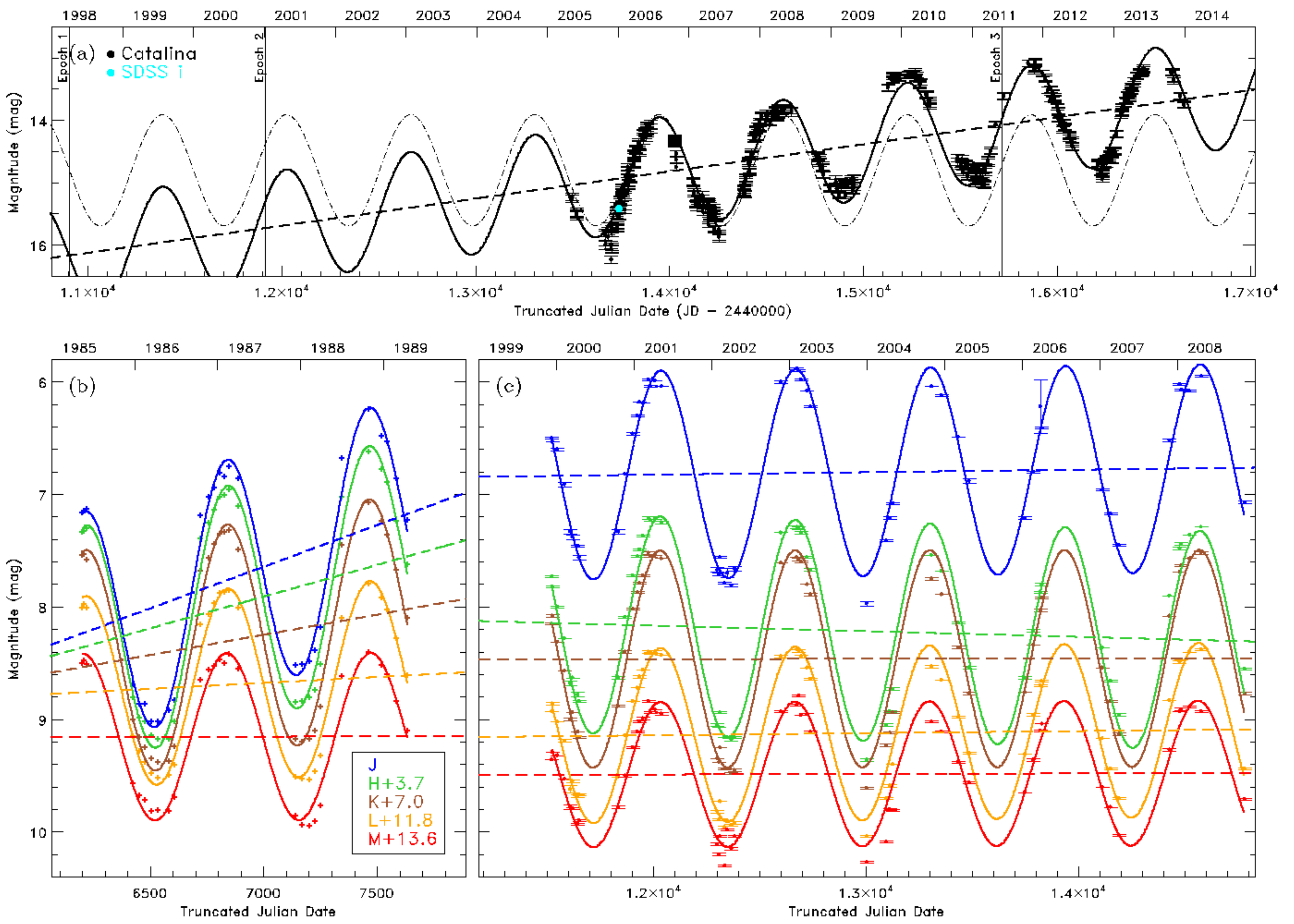}
  \caption{\label{fig:lc3}
    Optical and infrared lightcurves of IRC+10216, and fitted sinusoids. 
    (a) Top panel: Catalina lightcurve for 9 years (2005--2013). One SDSS 
    $i$ band data point (Dec. 2005) is plotted as a cyan circle (see text 
    for detail). The fit (solid curve) consists of a sinusoidal term with 
    a period of $640\pm0.9$ days and an amplitude of $1.8\pm0.02$ mag 
    (dot-dashed), and a linear term for brightening trend of $-0.16\pm0.004$ 
    mag yr$^{-1}$ (dashed). Vertical lines indicate the three relevant 
    \emph{HST} epochs in Figure\,\ref{fig:hst}.
    (b) Bottom left panel: $JHKL'M$ lightcurves of \citet{leb92} spanning 
    4 years (1985--1989). Arbitrary vertical shifts were used for comparison. 
    The average pulsation period of best fits is $627\pm26$ days, which 
    is slightly, but not significantly, smaller than the model ignoring 
    the linear term \citep[649 days,][]{leb92}. The $J$, $H$, and $K$ 
    lightcurves show systematic brightening trends at rates of $-0.3$, 
    $-0.2$, and $-0.1$ mag yr$^{-1}$, respectively, with the 
    uncertainties of 0.2 mag yr$^{-1}$.
    (c) Bottom right panel: $JHKLM$ lightcurves of \citet{she11} for 9 years 
    (2000--2008). The line colors for bands are same as in (b). The average 
    amplitude ($1.7\pm0.2$ mag) and period ($633\pm6$ days) are similar to 
    the ones in (b), but the brightening trends in near-infrared are not 
    found. The pulsation phase matches the Catalina ephemeris. 
  }
\end{figure*}

One of the exciting findings from the 2011 \emph{HST} image is the emergence 
of a point-like source at 0\farcs5 east of the central star (indicated by an 
arrow in Figure\,\ref{fig:hst}(d)). This source is projected within the dust 
lane of \citet{jef14}. The FWHM of its profile, 0\farcs15, is within the 
range of 0\farcs09--0\farcs17 measured for 10 unsaturated images of isolated 
stars. The color ($I_{\rm F814W}/I_{\rm F606W}$) of this point-like source 
is redder than those of diffuse clumps scattered over the inner $1\arcsec$ 
region (denoted by 1 to 4 in Figure\,\ref{fig:hst}(c)) by a factor of 
$2.9\pm1.1$, suggesting a different nature from the ambient matter. 
We posit that this point-like source is a companion star of IRC+10216. 

Its $r$ magnitude determined by referencing to SDSS stars is 
$r=21.1\pm0.2$ mag. Assuming that circumstellar extinction follows 
the interstellar extinction law and adopting the distance 130\,pc, 
one can derive physical parameters of this companion candidate. For 
a $r^{-2}$ density distribution of the envelope, the column density to 
the source at radius $R$ is calculated to be $0.5\pi R_*/R$ of that to 
the center, where $R_*$ represents the radius of dust formation zone 
(15\,AU is adopted). For a visual optical depth $\tau_{(0.55\mu m)}=25$ 
derived from an SED fitting by \citet{lad10}, the visual extinction 
toward the companion candidate would be $A_V\sim8$\,mag. With this 
extinction, its $r$ magnitude best matches a main-sequence star with 
a spectral type of M1 ($\sim0.5\,\Msun$). It is, however, very tentative 
because of uncertainties in the circumstellar extinction coefficients, 
inhomogeneity of AGB wind, and unknown line-of-sight location of the 
point-like source. 

If the point-like source seen in 2011 is indeed a binary companion, 
we can assess its possible orbit and proper motion. Assuming that 
the projected separation of 0\farcs5 (i.e., 65\,AU) corresponds to 
the orbital semi-major axis $a$, the orbital velocity of a binary 
component with respect to the other is $v=6\,\kmps(M/2.5\Msun)^{1/2}$, 
where $M$ is the total binary mass. This speed corresponds to 
0\farcs01 per year for an assumed total binary mass of $2.5\Msun$. 
Assuming a fitting accuracy of $\sim0\farcs02$, its movement may be 
observable in a few years from 2011 in the best case. The formula 
for binary dynamics $(p/{\rm yr})^2=(a/{\rm AU})^3(M/\Msun)^{-1}$ 
yields the orbital period $p$ of 330\,yr\,$(M/2.5\Msun)^{-1/2}$. 
This period is consistent with the timescale of 200--800\,yr derived 
from the intervals between the outer shell patterns \citep{cra87,mau99,
mau00}, which, in binary-induced spiral-shell models, represents the 
orbital period \citep[e.g.][]{kim12b}. 

The candidate for companion of IRC+10216 is proposed based on 
its point-like appearance, projection onto the dust lane, and color 
difference from nearby diffuse clumps. Its photometric and orbital 
period properties are not inconsistent with a low-mass main-sequence 
star. However, the absence of the companion candidate in the earlier 
epochs remains an open question since its orbital motion, if 
\emph{circular}, is rather slow to explain its relative position. 
A speed $>30\,\kmps$ is required to move 0\farcs5 in 10 years. 
From a single imaging data, we cannot rule out a possibility that 
this point-like source could be a \emph{compact} clump originated 
from small scale fluctuations in the dust condensation radius of 
IRC+10216. Further observations and quantitative modeling are 
desired to better investigate this object in the current and near 
future, especially since the giant star in IRC+10216 will reach the 
next light minimum in March 2016 when the companion is less overwhelmed.

\acknowledgments
This paper is dedicated to the memory of Patrick J. Huggins. 
We thank the anonymous referee, Ronald E. Taam, Thibaut Le Bertre, Andreas 
Mayer, Paul Berlioz-Arthaud, Hiroyuki Hirashita, and Noam Soker for their 
helpful suggestions to improve this paper. 
H.K. acknowledges an East Asian Core Observatories Association 
Fellowship.
This study used the NASA/ESA \emph{Hubble Space Telescope} observations, 
obtained from the data archive at the Space Telescope Science Institute. 
The CRTS survey is supported by the U.S.~National Science Foundation 
under grants AST-0909182 and AST-1313422.

\end{document}